\documentclass{hep99}
\begin{document}


\begin{titlepage}

\begin{flushright}
IRB-TH-6/99\\
August, 1999
\end{flushright}

\vspace{2.0cm}

\begin{center}
\Large\bf On the perturbative approach to the penguin-induced \boldmath $B \rightarrow \pi \phi$ 
\unboldmath decay \\
\vspace*{0.3truecm}
\end{center}

\vspace{1.8cm}

\begin{center}
\large Bla\v zenka Meli\'c\dag\\
{\sl Theoretical Physics Division, Rudjer Bo\v skovi\'c Institute,
P.O.Box 1016, HR-10001 Zagreb, Croatia\\[3pt]
E-mail: {\tt melic@thphys.irb.hr}}
\end{center}

\vspace{1.5cm}

\begin{center}
{\bf Abstract}\\[0.3cm]
\parbox{13cm}
{Using a modified perturbative approach that includes the Sudakov resummation
and transverse degrees of freedom we analyze the penguin-induced $B^{-}
\rightarrow \pi^{-}\phi$ decay.
The perturbative method enables
us to include nonfactorizable
contributions and to control virtual momenta appearing in the process.
The calculation supports the results obtained in the standard BSW factorization
approach, illustrating the electroweak penguin dominance and the
branching ratio of order ${\cal O}(10^{-8})$.
However, the estimated prediction of $16\%$ for CP asymmetry is much
larger than that obtained in the factorization approach.
}
\end{center}

\vspace{2.0cm}

\begin{center}
{\sl Talk given at the\\
International Europhysics Conference on High Energy Physics --
EPS-HEP '99,\\
Tampere, Finland, 15--21 July 1999\\
To appear in the Proceedings}
\end{center}

\fntext{\dag}{Supported by the Ministry of Science and Technology of the 
Republic of Croatia under Contract No. 00980102.}
\end{titlepage}
\thispagestyle{empty}
\vbox{}
\newpage

\setcounter{page}{0}


\title{On the perturbative approach to the penguin-induced $B \rightarrow \pi \phi$ decay}

\author{Bla\v zenka Meli\'c}
%

\address{ Theoretical Physics Division, Rudjer Bo\v skovi\'c Institute, 
P.O.Box 1016, HR-10001 Zagreb, Croatia\\[3pt]
E-mail: {\tt melic@thphys.irb.hr}}

\abstract{
Using a modified perturbative approach that includes the Sudakov resummation
and transverse degrees of freedom we analyze the penguin-induced $B^{-}
\rightarrow \pi^{-}\phi$ decay. 
The perturbative method enables
us to include nonfactorizable
contributions and to control virtual momenta appearing in the process.
The calculation supports the results obtained in the standard BSW factorization
approach, illustrating the electroweak penguin dominance and the
branching ratio of order ${\cal O}(10^{-8})$.
However, the estimated prediction of $16\%$ for CP asymmetry is much
larger than that obtained in the factorization approach.
} 

\maketitle


\section{Introduction\label{sec:intro}}

Nowadays, experimental facilities offer a possibility of searching for CP asymmetries
in penguin-induced nonleptonic decays, very promising decays to detect
direct CP violation. Such decays have small branching ratios (BR), but satisfy
both requirements for CP-violating asymmetry, owing to the fact that
penguins are loop diagrams with different quark generations contributing with
different weak CP-phases from the CKM matrix and that
final-state strong interaction phases emerge from the  absorptive part of
penguin amplitudes.

Our aim is to investigate the penguin-induced $B^{-} \rightarrow \pi^{-}\phi$ 
decay in the modified perturbative approach.


We present a complete calculation of factorizable
and nonfactorizable contributions in the 
$B^{-} \rightarrow \pi^{-}\phi$ decay up to order $\cal O (\alpha_{\rm s}
\alpha_{\rm em})$. 
Nonfactorizable contributions from the QCD-penguin operators are examined and
their role in the EW-penguin dominated processes, such as
$B^{-} \rightarrow \pi^{-}\phi$, is assigned. Direct CP-asymmetry violation is 
calculated and all results are compared  
with those obtained in the standard BSW factorization approach.

\section{Perturbative model\label{sec:pert}}

The nonleptonic $B^{-} \rightarrow \pi^{-}\phi$ decay is governed by the weak
decay of the heavy b-quark, $b \rightarrow d s \overline{s}$. The light
antiquark of the $B$ meson is the spectator in the decay, being only slightly
accelerated by the exchange of a hard gluon to form a pion in the final state.

We use the NLO weak Hamiltonian in which the renormalization-scheme
dependence
of the Wilson coefficients is explicitly canceled by the inclusion of the
one-loop QED matrix elements of the tree-level operators
${\cal O}_{1,2}^{(q)}$. 
The final expression for the matrix element is then
\begin{eqnarray}
\lefteqn {\langle {\pi^{-}} \phi | {\cal H}_{eff} (\Delta = -1) | B^{-} \rangle =
\frac{G_F}{\sqrt{2}} \sum_{q = u,c} V_q }
\nonumber \\
& & \hspace{-0.5cm}\times \left ( \sum_{k=3}^{10}
\overline{c}_k(\mu) \langle {\cal O}_k \rangle^{\rm tree}  
+ \frac{\alpha_{\rm em}(\mu)}{9 \pi}
( 3 \overline{c}_1(\mu) + \overline{c}_2(\mu))\times
\right . \nonumber \\
& &  \hspace{-0.5cm}\left .
\left \langle  \left ( \frac{10}{9} -
\Delta G(m_q^2, q^2, \mu^2) \right )
( {\cal O}_7 + {\cal O}_9 )
\right \rangle^{\rm tree} 
\right )
\label{eq:ME}
\end{eqnarray}
and here $\langle {\cal O}_k\rangle^{\rm tree} \simeq \langle \pi^{-}\phi |
{\cal O}_k | B^{-} \rangle$. 

The absorptive part of the matrix element needed for a nonvanishing CP asymmetry 
resides in the 
$\Delta G(m_q, q^2, \mu)$ function for $q^2 \ge 4 m_q$. In the perturbative 
calculation, the process-dependent 
virtual momentum $q^2$ is determined by the momentum distribution in the decay and 
it is controlled by the momentum distributions in the process.

The nontrivial part is the calculation of the matrix elements of the 
four-quark operators $\langle {\cal O}_k \rangle^{\rm tree}$ at the tree level, 
which we are going to perform in the modified 
perturbative approach \cite{LS}.

The matrix element factorizes into the
convolution of distribution amplitudes of hadrons involved into the decay
(hadron wave functions) and the hard scattering amplitude of valence partons. 
The hard scattering amplitude can be straighforwardly calculated perturbatively, 
taking into
account all possible exchanges of a hard gluon between valence partons in a
given $\alpha_s$-order of the calculation \cite{JA}.

On the other hand, the hadronic wave functions represent the
most speculative part of the perturbative approach. They are of nonperturbative
 origin and should be a universal, process-independent quantity. However, there 
are several models of wave functions for each of the particles and it is not 
simple to rule some of them out.

We make a selection among the wave functions by comparing the results for the 
$ B \rightarrow \pi$ transition form factor obtained using nonperturbative methods
(lattice, QCD sum rules) with those estimated in our modified perturbative approach. 

The selected wave functions for the B-meson, the pion, and the $\phi$-meson are,  
respectively, 
\begin{eqnarray}
\lefteqn {\Phi_B(x) \propto
\sqrt{x (1-x)}\; \exp\left
(- \frac{M_B^2}{2 \omega^2} x^2 \right ) ,}\\
& & \hspace{-0.65cm}\Phi_{\pi}^{CZ}(x,\mu_1) \propto 6 x(1-x) \nonumber \\ 
& & \times \left (1 + (5 (1-2 x)^2 -1) 
\left ( \frac{\alpha_s(\mu_1)}{\alpha_s(\mu_{0})} \right )^{50/81} 
\right ), \\
& &  \hspace{-0.65cm} \Phi_{\phi}(x) \propto 6 x(1-x).  
\end{eqnarray}
%
Having selected these wave functions, we are now in a position to 
give some predictions. For details, the reader is referred to \cite{JA}.

\section{Numerical results and conclusions\label{sec:results}}
The results are presented in Tables I and II. 
The results in column II are obtained by the calculation, 
in which the $\mu$
scale-setting ambiguity of
Wilson coefficients is moderated by applying the three-scale
factorization theorem \cite{LiColl}. 
The theorem keeps trace of all three scales characterizing the
nonleptonic weak decay: the W-boson mass $M_W$, the
typical scale $t$ of the process, and the hadronic scale $\sim \Lambda_{QCD}$,
and proves for the leading-order weak Hamiltonian that Wilson coefficients
should be taken at the scale $t$. 
The matrix elements of the operators ${\cal O}_k$ and
Wilson coefficients are then both calculated at the same scale,
contrary to the standard calculation with the Wilson coefficients 
${\overline c}_k(\mu = m_b)$ represented in column I.

Our results for the branching ratio appear to be in agreement
with previous calculations performed
in the BSW factorization approach \cite{Fleischer}, 
predicting the branching ratio to be of
order ${\cal O}(10^{-8})$, dominated by the EW penguins. On the other hand,
the predicted CP asymmetry differs a lot from that estimated in
the BSW factorization approach, being as large as $16\%$ and having
an opposite sign for the
preferred values of the CKM parameters $\overline{\rho} = 0.16$ and
$\overline{\eta} = 0.33$. The large
CP asymmetry estimated in the perturbative approach is the result of
large on-shell effects of the virtual propagators involved in the 
calculation \cite{JA}.

Besides, if the Wilson
coefficients are considered to be functions of the scale (results in column II),
the same one
which appears in the hadronic matrix elements, then the nonfactorizable
QCD penguin contributions appear to be negligible, as is the case with
the obviously very small nonfactorizable
contributions of the EW penguin operators.
Furthermore, estimations based on this assumption 
have produced the branching ratios about
a factor of two larger than those calculated
with the conventional Wilson coefficients.


\begin{table}\caption{ Branching ratios 
calculated for different
penguin contributions, by taking only the factorizable parts or the complete
expression into account (see text). 
The CKM  parameters used are $\overline{\rho} = 0.16$ and 
$\overline{\eta} = 0.33$.}
\begin{tabular}{lccc}
\hline 
Penguin contributions &
\multicolumn{3}{c}{BR/$10^{-10}$}\\  \cline{2-4} & BSW & I & II \\ \hline
\\
$\rm QCD^{fact}$ & $0.20$ & $0.14$
& $1.06$ \\
$\rm QCD^{all}$ & -& $2.51$  & $0.73$ \\
\\
$\rm (QCD+QED)^{fact}$ & $34$ & $38$ & $89$ \\
$\rm (QCD+QED)^{all}$  &- & $44$  & $85$ \\
\\
\hline 
\end{tabular}
\label{t:rBR}
\end{table}

\begin{table}
\caption{ CP asymmetries 
calculated for the QCD and QED penguin contributions together,
by taking only the factorizable parts or the complete
expression into account. }
\begin{tabular}{lccc}
\hline 
Penguin contributions&
\multicolumn{3}{c}{$a_{CP}/10^{-2}$} \\
  \cline{2-4} &
BSW & I & II \\ \hline
\\
$\rm (QCD+QED)^{fact}$  & -1.9 & 14.6 & 15.4  \\
$\rm (QCD+QED)^{all}$  &  - & 16.1 & 16.3  \\
\\
\hline 
\end{tabular}
\label{t:rCP}
\end{table}

\end{document}